\documentclass[twocolumn]{aastex62}

\usepackage{url}

\usepackage{amsmath}
\usepackage{rotating}

\graphicspath{{./}{figures/}}

\newcommand{\R}{{\mathbb R}}

\shorttitle{Detecting TNOs with TESS }
\shortauthors{Ganapathi}

\begin{document}

\title{Maximum Likelihood Systematic Effect Modeling and Matched Filtering to Detect Trans-Neptunian Objects with TESS}

\author{Varun Ganapathi}
\affiliation{AKASA, Inc., South San Francisco, CA 94131, USA}

\correspondingauthor{Varun Ganapathi}
\email{varung@cs.stanford.edu}

\begin{abstract}

We present a pipeline for searching for trans-Neptunian objects (TNOs) using data from the TESS mission, that includes a novel optimization-based framework for subtracting the effects of scattered light and pointing jitter. The background subtraction procedure we adopt, when combined with a moving average, allows one to see TNOs such as 90366 SEDNA, 2015 BP519 with the “naked eye.” Moreover, this procedure also enabled us to identify two TNO candidates via direct visual observation (subsequently identified to be 2003 UZ413 and 2005 RR43). To automate the extraction of candidate TNOs, we apply a matched filter that can be tuned to objects at different distances and orbital inclinations. We also demonstrate the performance of the algorithm by recovering signals of three trans-Neptunian objects automatically with a high level of confidence. We further validate the approach via synthetic experiments that test recovery rate as a function of magnitude and distance. We find that there is a trade-off between distance and magnitude; controlling for magnitude, it is easier to detect faster moving objects. Our method can detect objects at distances of 250 AU for magnitudes of +21, and closer objects at fainter magnitudes. On a single contemporary GPU (NVIDIA A100) the method can search for 100 trajectories on 1000 2048 x 2048 frames in 5 minutes, dramatically faster than previous approaches. This method can be used to perform large scale fully automated surveys for TNOs and potential Planet Nine candidates. 

\end{abstract}

\keywords{Trans-Neptunian objects (1705) -- detached objects (376) -- minor planets (1065) -- solar system (1528) -- planetary theory (1258) -- sky surveys (1464) -- Planet Nine (?)}

\section{Introduction} 
\label{section:intro}

The confluence of space telescopes, cloud computing and hardware accelerators enables a new era of astronomy, especially for citizen science. The Transiting Exoplanet Survey Satellite (TESS) \citep{ricker2015tess} is a space telescope that is designed to detect transiting exoplanets. The full-frame images that it captures are transmitted to earth after which they are processed and uploaded into cloud storage ``buckets'' hosted on Amazon Web Services (S3).  There they are able to be downloaded at high speed into any cloud computing provider.
Cloud computing providers offer relatively inexpensive on-demand access to virtual machines with GPUs.

While TESS was designed to capture transiting exoplanets, it can also be leveraged to find large objects in our own solar system. Our goal in this paper is to develop an algorithm to find candidate TNOs that can be confirmed via observation with other telescopes. There has been much previous work in this direction using other sources, such as the Dark Energy Survey \citep[DES;][]{dark2005dark} and indirect gravitational searches \citep{fienga2016, holman2016observationalc, holman2016observationalp}. 

Recently a proof-of-concept survey was performed using TESS data itself \citep{rice2020}, and is based on the basic idea of shift-stacking, which has also been called ``digital tracking'', ``de-orbiting'' or ``pencil-beam surveys.'' These methods have also been applied to find solar system satellites\citep{holman2004discovery, kavelaars2004discovery, burkhart2016deep} and TNOs \citep{gladman1998pencil, gladman2001structure, bernstein2004size}. The reported compute time of the shift-stacking implementation in \citep{rice2020} is 7-8 hours per $256 \times 256$ frame. In contrast, the algorithm presented in this paper can process a $2048 \times 2048$ frame in about 3-5 minutes (not including the 1 minute of data download time, depending on search parameters), which is thousands of times faster.

We attempt to improve upon previous approaches by developing a framework that explicitly models scattered light and sub-pixel pointing jitter, followed by a matched filter which incorporates the known pixel response function. We avoid the use of PCA (or co-trending basis vectors \citep{Kinemuchi_2012}. Matched filtering is a provably optimal maximum likelihood based technique \citep{rothstein1954probability} for identifying a known signal corrupted with additive white noise, and has been applied to a variety of applications, including gravitational wave identification \citep{owen1999matched}. The algorithms we chose are efficient and amenable to computation on GPUs which allows ours method to run many times faster than previous approaches. To facilitate easy comparison with the shift-stacking method, we follow \citep{rice2020} and evaluate our method on Sectors 18 and 19, which lie directly along the galactic plane.

\section{Data Overview}
\label{section:data_overview}
 The TESS mission is designed to survey the sky in 13 sectors, each of which is 27 days long.  As a result, over the course of a year, TESS will have observed each location in a hemisphere for 27 days. There are 4 cameras stacked in column, each consisting of 4 CCDs that have a resolution of 2048 x 2048. TESS is sensitive to stars with $V \lesssim 20$ and captures full-frame images (FFIs) integrated over multiple minutes. During the initial two years each FFI represented a numerical sum of 30 minutes of data, while in the next two years, the cadence decreased to 10 minutes. Each pixel spans $21\arcsec\times21\arcsec$ for a combined total field of view $24\degr\times96\degr$.  

We download all frames, which have been corrected for instrument/detector artifacts, directly from the Mikulski Archive for Space Telescopes (MAST) and from the public S3 bucket made available by MAST.\footnote{\url{http://archive.stsci.edu/tess/bulk_downloads/bulk_downloads_ffi-tp-lc-dv.html}}.

TESS points roughly away from the sun towards a fixed point in the sky for 27 days at a time. Since TESS orbits around the Earth, TESS itself is moving roughly perpendicularly to the direction it is viewing. During each sector, TESS is locked onto a given reference point (RA, Dec). The orbital motion of TESS throughout the sector creates a baseline that causes objects to appear to move (with velocity inversely proportional to distance) due to parallax.  Objects that are closer will move more, both due to their own true velocity in space (again inversely proportion to distance), and due to the motion of the viewer (TESS). The range of distances we are interested in (50AU to 800AU), produce a parallax displacement of 80 pixels to 5 pixels in 27 days. For most sectors in the survey, the rows of the TESS camera are aligned approximately with the ecliptic plane, thus the parallax motion will be along a row of pixels. Depending on an objects distance and orbital inclination, we can calculate its trajectory in pixel space as $\frac{4000px}{dist(AU)}$. A detailed derivation can be found in \citep{rice2020}

The TESS pixel response function (PRF) determines the observed appearance of point light sources. Approximately 50\% of the flux due to a light source will be measured in the peak pixel, and 90\%  within a 2 x 2 square of pixels. We can compute the added flux due to a moving object of magnitude $m$, by convolution of the PRF with the expected trajectory and scaling by the flux for that magnitude. 

\subsection{Systematic Effects}
\label{section:systematic_effects}
\begin{figure}
    \centering
    \includegraphics[width=0.48\textwidth]{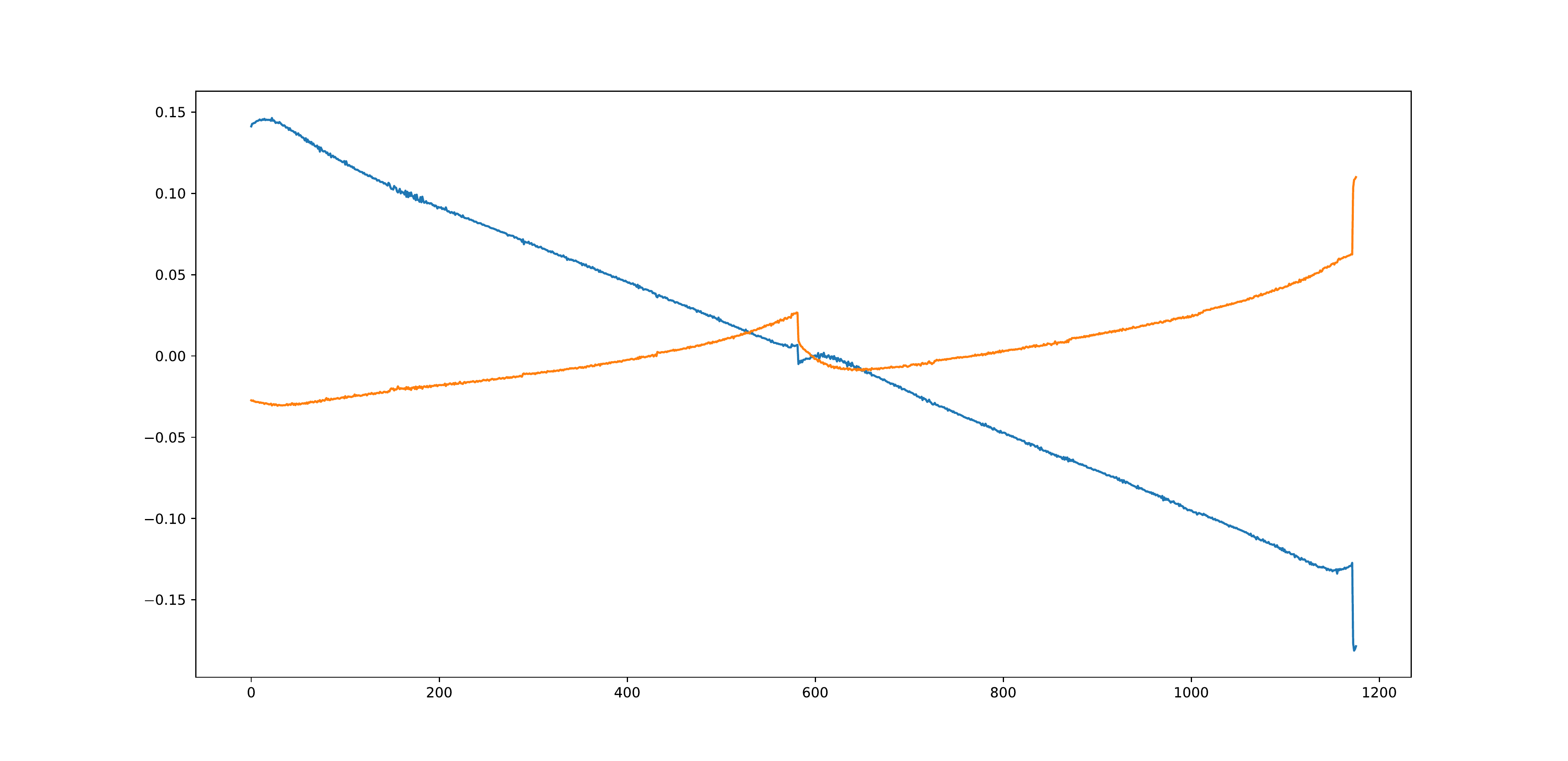}
    \caption{Pointing jitter in reference point pixel coordinates (row and column) vs time in Sector 5. X-axis is FFI index over the course of 27 days, and Y axis is the change in the location of the reference point in pixels.}
    \label{fig:jitter}
\end{figure}

The TESS spacecraft exhibits a small amount of positional jitter in its pointing direction. This jitter is generally less than 1 pixel. Figure~\ref{fig:jitter} shows the changing pixel location of the reference point in Sector 5, which is at up to $\pm .15$  pixels in this example. The jitter in the example was produced by taking the center reference point in (RA, Dec) on the first frame and computing its pixel location on each frame. We define jitter as the change in the projected location over time. TESS accurately estimates its own pointing direction by tracking a large collection of bright stars, which it uses to drive its attitude control system. The reason jitter is important, despite how small it is, is that the PRF is sensitive to exact location of the light source. The measured brightness of various pixels will vary as a star moves across CCD pixels. For instance, if a light source is directly in the middle of a single pixel, that pixel will contain most of its charge vs if the light source is at the corner of four pixels. In the latter case the charge will be split equally into 4 pixels rather than mostly distributed into the one center pixel. As the spacecraft jitters, the pixels slide across the light sources causing a varying brightness. It is this jitter than causes the appearance of ``dipoles'' when one subtracts the median frame from all the frames.

\begin{figure}
    \centering
    \includegraphics[width=0.48\textwidth]{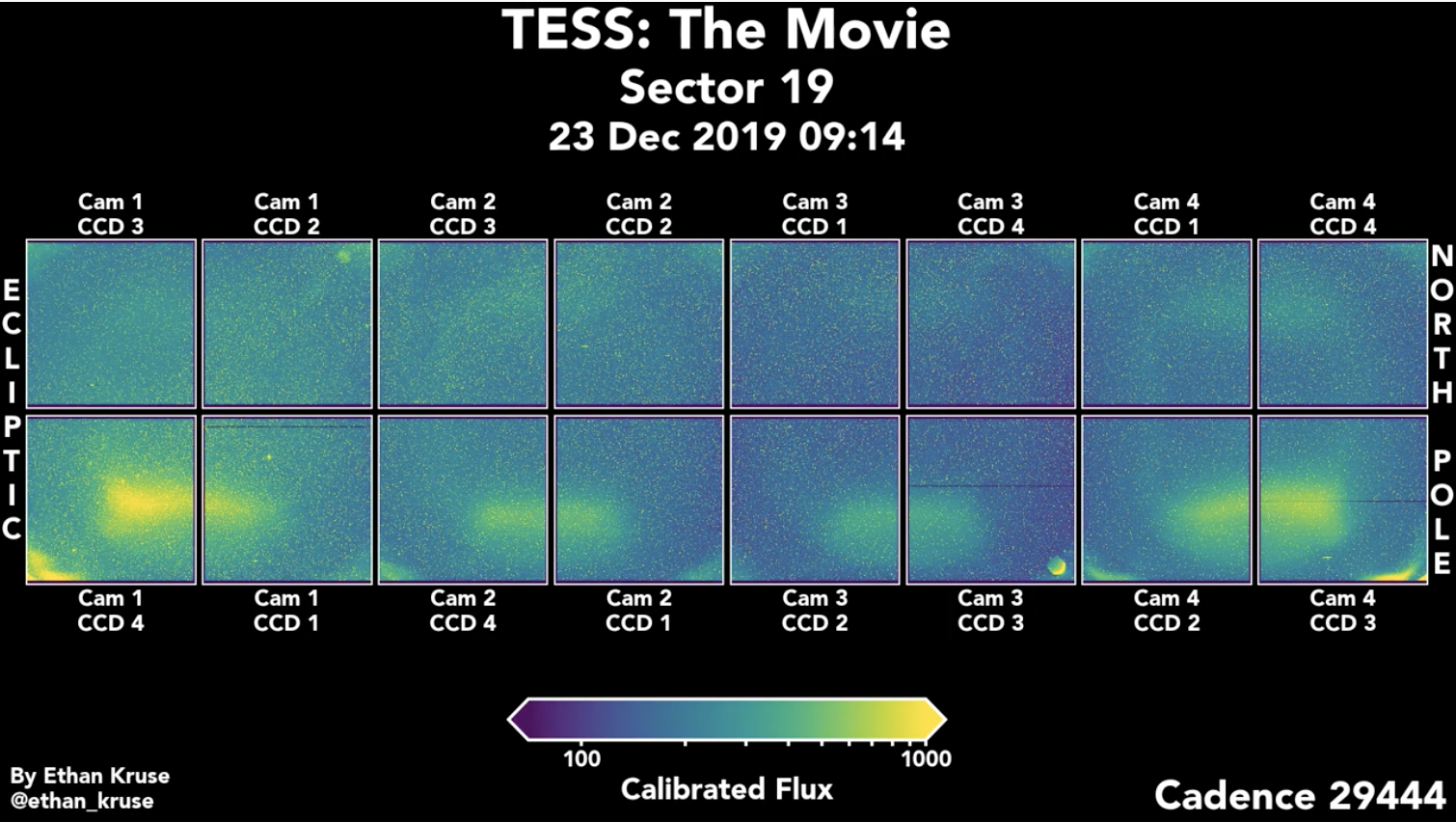}
    \caption{Scattered light in Sector 19. Source \url{https://heasarc.gsfc.nasa.gov/docs/tess/observing-technical.html}}
    \label{fig:scattered_light}
\end{figure}

TESS also is affected by scattered light from the Earth and moon(Figure~\ref{fig:scattered_light}). Scattered light appears to vary slowly with time and in space, and is due to external unintended light entering the lens. This effect is very spatially correlated and smoothly varying over time.

\section{Methods}
\label{section:methods}
At a high level, our approach is to define hidden variables for the various effects we seek to model. Then we optimize the values for the hidden variables to maximize the likelihood of the observed data. We construct a forward photometric model of the expected flux as a function of assumed-known camera orientation and hidden variables which are the stellar background, the scattered light and the estimated parameters of moving objects.

\subsection{Forward Model}
For a fixed moment in time $t$, let $\theta_t$ be the jitter (the difference of the pixel location of reference point at time $t$ to its pixel position in the reference frame). $\theta_t = (x_t, y_t)$. Let $\theta_t^k = (x_t, y_t, x_t^2, y_t^2, x_t y_t)$. The purpose of $\theta_t^k$ is to provide a basis for a Taylor expansion.

We first solve for scattered light at low-resolution, and then solve for the Taylor expanded background model at high resolution. Let $N = 512$, and $n = 64$. Let $B \in R^{N \times N}$ be the background flux, $B_{\theta} \in R^{N \times N \times k}$ be the parameters for a 2nd order spatial Taylor expansion of the background image, and $S \in R^{n \times n}$ be the low-resolution scattered light image, $M$ be the moving object flux, $D$ be the data downsampled to $512 \times 512$ from $2048 \times 2048$, and $P$ be the pixel response function. $U$ is a function that upsamples $S$ from $n$ to $N$. In our implementation, $U$ performs bilinear interpolation to upsample a $n \times n$ image to a $N \times N$ image. Upsampling by bilinear interpolation simply means that the high resolution pixel is weighted combination of the 4 nearest low resolution points.

The photometric model is defined as follows:
\label{section:forward_model}
\begin{equation}
    \label{eq:forward_model}
    D_t = U(S_t) + B + B_\theta \theta_t^k + \epsilon_t
\end{equation}

This model says that the data at time $t$ is the scattered light at time $t$ upsampled to high resolution, the background, the background interpolated to the jittered position, and some noise $\epsilon$. This lets us represent each pixel in B as a quadratic surface. Again, this is a linear function because the $\theta_t$ are known observed constants at each point in time. We are simply constructing basis functions from them and weighting those functions by learned parameters.

Given the forward model, we can construct a negative log likelihood function which minimize. 
\begin{align}
    \label{eq:loss_function}
    L &= \sum_{t,i,j} (D_{t,i,j} - \hat{D}_{t,i,j})^2 \\
      &= \sum_{t,i,j} L_\delta(D_{t,i,j} - (U(S_t) + B + B_\theta \theta_t^k))_{i,j})
\end{align}

The $L_\delta(a)$ function is a smooth approximation to the absolute value function, often used to reduce the effect of outliers compared to a squared error. It is defined as: 
\begin{equation}
  \label{eq:huber}
  L_\delta(a) = \begin{cases}
    \frac{1}{2} a^2 & \text{if } |a| \leq \delta_0 \\
    \delta_0(|a - \frac{1}{2}\delta_0) & \text{otherwise}
\end{cases}
\end{equation}
 
We minimize this convex loss over $B,B_\theta,S_t$ using the L-BFGS \citep{zhu1995limited} algorithm.

Since the function is convex, the algorithm will achieve a globally optimal solution regardless of the initial parameters. The output of the procedure is an estimate of the scattered light for every single image. If we were to solve this problem using the full resolution data, our computation would be complete. However, for computational efficiency, and since we are estimating the scattered light at low-resolution, we use a low-resolution version of the data to estimate the scattered light.

Once we solve for scattered light $S_t$ at low resolution, we hold it fixed, and re-estimate $B,B_\theta$ at high resolution using the same loss function. Since the scattered light $S_t$ is fixed, this separates into independent problems for each pixel in the background model $B_{i,j}, B_{\theta,i,j}$. Since we are effectively regressing each pixel independently against the same input data $\theta,\theta^k$, we can reuse the computation of inverse of the Gram matrix (or normal matrix) to solve each problem. Note that the loss function is not least squares, but the Huber robust loss function. Nonetheless, it can be shown that you can optimize the Huber loss through repeated least squares solves, alternated with clipping the residuals\citep{huber1977}. One distinguishing aspect of our approach is the explicit use of a spatial model of scattered light. Other approaches implicitly estimate scattered light as part of the per-pixel fitting procedure, which throws away the information that scattered light is spatially smooth.

\subsection{Matched Filter}
\begin{figure}[h]
    \centering
    \includegraphics[width=0.25\textwidth]{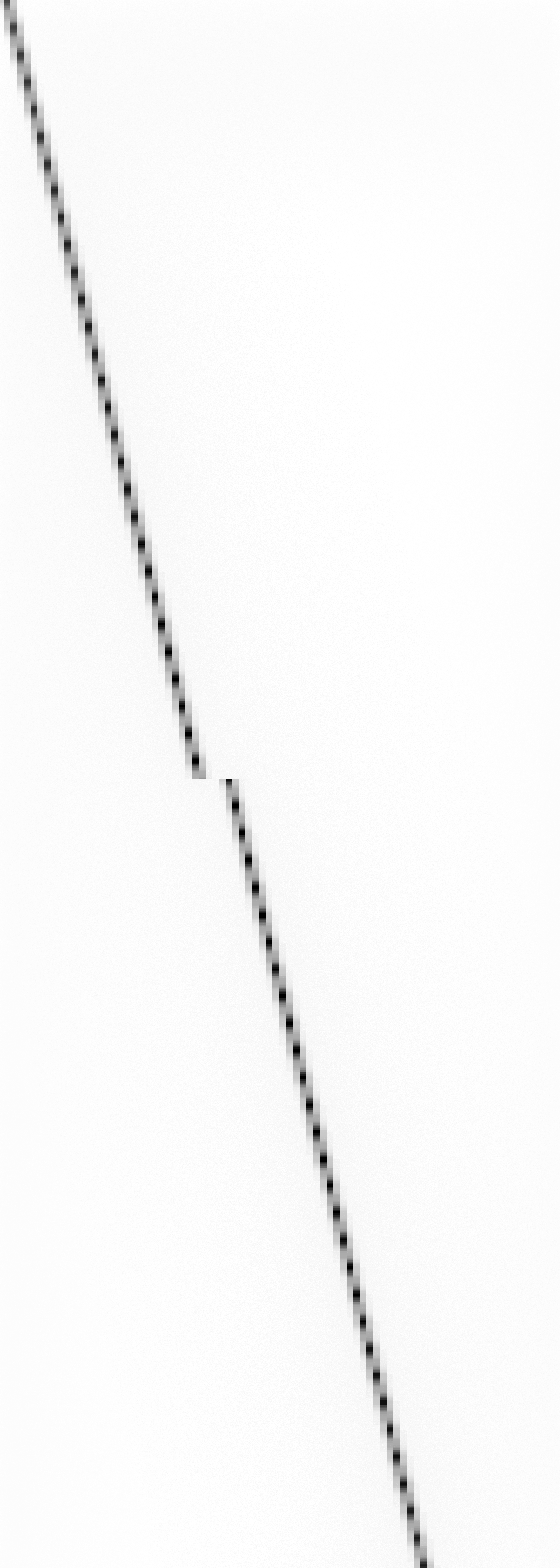}
    \caption{Example center row of matched filter for a purely horizontal translation. Each filter is a $T \times C \times R$ tensor. The first dimension is the thickness in rows, the second dimension is the number of columns, and the third dimension is the time. The discontinuity is due to the time-jump that occurs when TESS is transmitting frames to Earth.}
    \label{fig:patch}
\end{figure}

Figure \ref{fig:patch} shows an example matched filter. This is $F \in \R^{R \times C \times T}$ tensor that show the expected added flux due to a moving object corresponding to $dx=40, dy=0$. This filter is produced by convolving the line of the trajectory (including the pointing jitter) with the pixel response function. We can produce a set of such filters corresponding to each sample trajectory we want to test. Once we have subtracted the effects of scattered light and pointing jitter, what should be left is the sum of Poisson noise and the flux of moving objects. The standard deviation of Poisson noise is proportional to its mean, which we have estimated as the sum of the scattered light and background flux. We set the value of bright pixels to 0 because the noise in those pixels will so high to make it impossible to detect a magnitude 20+ signal. Clipping is very important because it helps remove the effects of unmodeled systematic effects such as fast asteroids, or time-varying bright stars. We approximate the noise on remaining pixels as a mean-zero Gaussian with constant standard deviation.

The matched filter is provably optimal for detecting a known signal in the presence of additive Gaussian noise. The process of applying a matched filter is taking the dot product of the filter with the underlying data signal. We can compute this dot product at all possible pixel locations via convolution. The result is a direct estimate of the total flux due to that moving object, which allows to not only estimate the position and trajectory of a detection, but also the magnitude.

\begin{equation}
    \label{eq:matched_filter}
    S = D \ast F
  \end{equation}

The resulting score is an estimate of the flux added by the moving object. As expected, brighter objects are easier to detect because their score is higher, while darker objects will have lower score, since it is harder to distinguish a dark object from background noise. By applying $N$ matched filters, we get a score for each of the $N$ evaluated trajectories. We take the maximum score across the filter bank, which then tells us the most likely trajectory, along with the score of that trajectory.

The shift-stack method in \cite{rice2020} can be viewed as an approximate matched filter. They use a single non-zero value, set to $1$, for each time-slice. The chosen non-zero value should be the maximum flux value in that time slice. This throws away approximately half of the signal in the data because the flux in the peak pixel is less than half of the total flux deposited on the sensor. Thus we might imagine that the error of our method is approximately $\sqrt(2)$ lower. There is a potential computational benefit of using such a sparse filter. However we compensate for a more accurate filter by leveraging highly optimized methods for convolution on GPUs. The computational complexity of the scoring is the product of the number of elements in the matched filters times the number of pixel locations tested. If this becomes a bottleneck, it could be sped up by at least an order of magnitude by representing the matched filter with a sparse matrix and using a sparse convolution CUDA kernel.

\subsection{Interpreting Scores}

The algorithm ultimately produces a score for every single pixel for the best matching filter. Each filter corresponds to a particular object trajectory which can be characterized by $dx$ and $dy$. Ultimately, we take all scores above a threshold as potential candidates.

Any choice of score threshold implies a recall rate and false discovery rate. The lower the threshold, the more potential candidates we will identify, but at the expense of increasing the number of false positives. To facilitate comparison with \citep{rice2020}, we compute the standard deviation $\sigma$ of the scores in the $255 \times 255$ square centered on a pixel. We represent our scores in units of $\sigma$. We consider candidates to be scores greater than $3 \sigma$, in accordance with their approach.

\section{Results}
\label{section:results}

\subsection{Injection Recovery}
\label{subsection:injection_recovery}
We performed a test of the method on synthetically generated data to characterize its performance. Since the matched filter itself corresponds to exactly the expected signal due to a moving object, signal injection corresponds to adding a filter times a flux to the raw data. We added a grid of moving objects spaced by 20 pixels vertically and 80 pixels horizontally (to ensure no overlap between injected signals). After injecting the data, we run the entire background subtraction process followed by the matched filter. The result is a set of candidates, and we can gauge the effectiveness of the method by the percent of known signals that were recovered at a particular significance level $3 \sigma$. To facilitate comparison with \citep{rice2020}, we evaluate the method on the $256\times256$ cutout centered in CCD 1 of each camera. We consider the effectiveness of the algorithm at recovering objects at various magnitudes and displacements in the $x$ direction. What we find is that the sensitivity of the method improves with magnitude and displacement. Far away objects, due to their small displacement across the sensor, are harder to detect. We surmise this is due to the fact that because they move slowly across the pixel, they can sometimes be removed by the background subtraction method. 

\begin{figure*}[ht]
  \centering
  \includegraphics[trim={0 5cm 0 0},clip,width=0.9\textwidth]{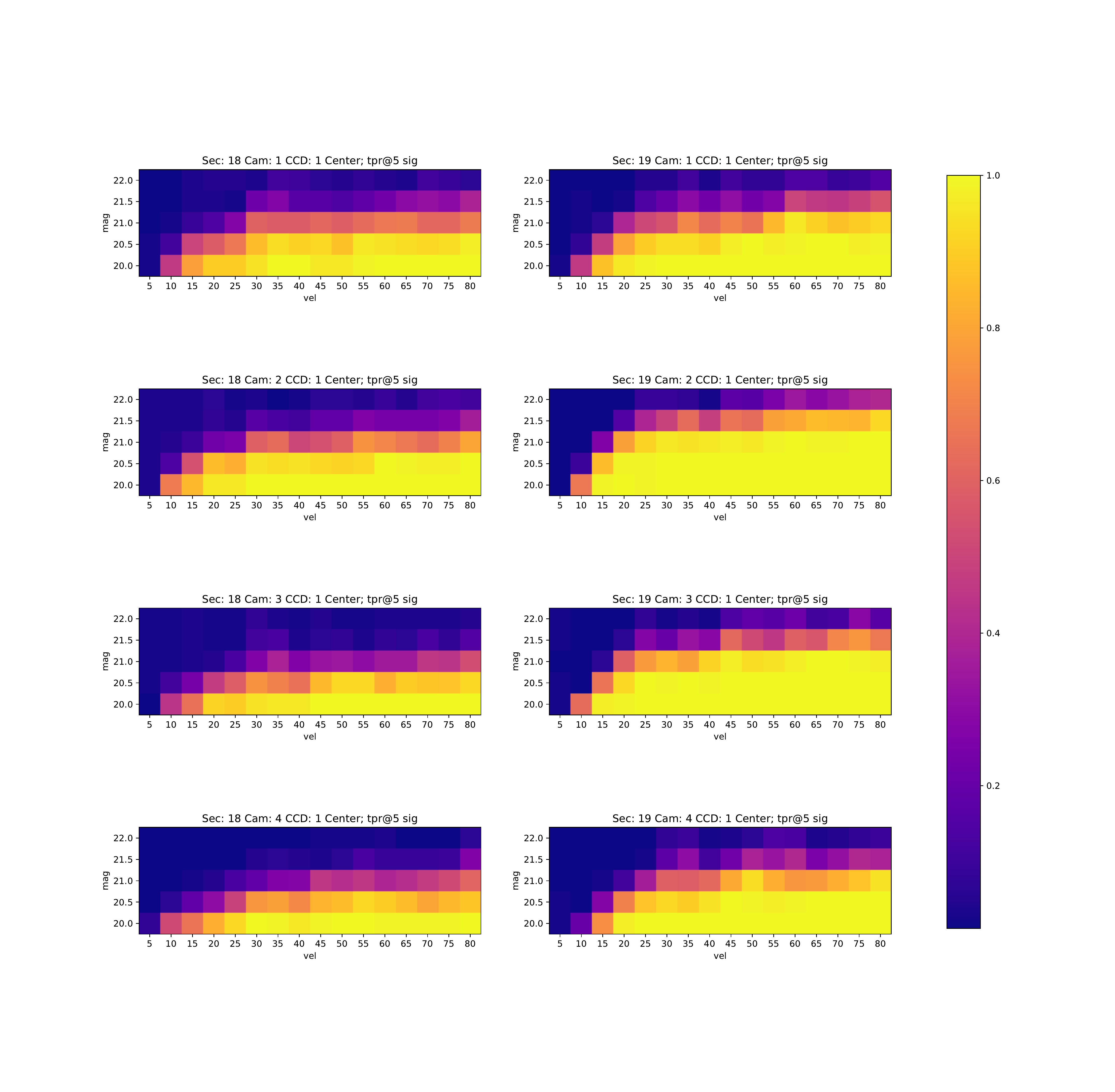}
\caption{Recovery rate vs magnitude and vel (displacement in pixels) for $5 \sigma$ detections }
\end{figure*}

Based on synthetic experiments, we see that the algorithm can pick up objects as faint as 22nd magnitude as far away as 160 AU. We can see objects as faint as 21.5 magnitude at a distance of 266 AU. However, the method requires objects to be much brighter to see them at further distances.

\subsection{Recovery of Known Objects}
\label{subsection:recovery_known_objects}
A good test is whether these methods can recover known outer solar system objects. We apply our method to sector 5, and show all the detected candidates, and cross-reference against three known TNOs in that sector. \citep{rice2020} provides a helpful table of three known TNOs. We have output the significance of the corresponding trajectory as scored by our algorithm.  We searched for objects that were at a distance between 35 and 800 AU ($dx=114$ to $dx=5$) with $dy=-6$ to $dy=0$.  Each trajectory we consider is a straight line, while some of trajectories are in fact highly non-linear. Nevertheless, we are able to recover these candidates with blind searches. In \citep{rice2020}, it appeared that 2007 TG422 was not found to have high confidence with the blind recovery approach, while our algorithm does find it with high significance ($\ge 6 \sigma$). 

We found that we can apply a simple moving average filter to the intermediate cleaned data and see candidate TNOs directly with the naked eye. The box filter averages frames in groups of $K$, where $K$ should be set to be less than the number of frames the target object will take to transit through one pixel.  When we view these frames in a loop, we can directly see Sedna and 2015 BP 519.  We also visually observed two objects which have high significance, and we subsequently identified them in the minor planet catalog. We have included these observations and their significance according to our algorithm in Table \ref{tab:detections}.  The matched filter can find faint objects that are hard to see via the moving average filter, and allows fully automatic candidate extraction.

\begin{deluxetable*}{ccccccccccc}
  \tablecaption{Recovery results for the three known TNOs visible in Sector 5.\label{tab:detections}}
\tablewidth{700pt}
\tabletypesize{\scriptsize}
\tablehead{
\colhead{Name} & \colhead{(S, Cam, CCD)} & \colhead{Type} & \colhead{y} & \colhead{x} & \colhead{dy} & \colhead{dx} & \colhead{$V$} & \colhead{$d$ (AU)} & \colhead{Significance} & \colhead{Visible in MAVG}}
\startdata
90377 Sedna & (5, 1, 4) & nominal & 1,101 & 1,497 & -1 & 46 & 20.86 & 86.2 & - & \\
 &  & algorithm & 1,102 & 1,497 & -1 & 48 & - & - & 30.90 $\sigma$ & Yes \\ \hline
2015 BP519 & (5, 3, 2) & nominal & 146 & 899 & -6 & 75 & 21.81 & 54.4 & - &  \\ 
 &  & algorithm & 148 & 899 & -4 & 77 & 21.64 & - & 10.36 $\sigma$ & Yes \\ \hline
2007 TG422 & (5, 1, 3) & nominal & 1,306 & 1,384 & -6 & 94 & 22.31 & 36.8 & - &  \\ 
 &  & algorithm & 1,306 & 1,385 & -5 & 94 & 22.16 & - & 6.17 $\sigma$ & No \\ \hline
 2003 UZ413 & (5, 1, 4) & nominal & 1129 & 1771 & -3 & 92 & - & - & - & - \\ 
 & & algorithm & 1130 & 1776 & -3 & 89 & 21.41 & - & 12.09 $\sigma$ & Yes \\ \hline
 2005 RR43 & (5, 1, 3) & nominal & 686 & 1693 & -9 & 98 & - & - & - & - \\ 
 & & algorithm & 687 & 1697 & -7 & 94 & 21.12 & - & 18.06 $\sigma$ & Yes \\
\hline
\enddata
\end{deluxetable*}

\section{Discussion}
\label{section:discussion}
The approach we've presented directly models systematic noise in the TESS data.
With a moving average filter, one can directly identify moving objects by eye,
and that the matched filter will find these same objects with high confidence.
The matched filter of course can find fainter objects that are hard to see
directly. We've provided a few examples of manually identified objects in
Sector 5, though we stress this was not a comprehensive search, but simply
chance discoveries. There are likely to be more objects that are visible that
we have not included in this paper. 

The background subtraction process runs very quickly on modern GPUs. It takes just a minute to process an entire sector's worth of frames for a single $2048 \times 2048$ CCD, which is split to about $40s$ to solve for the scattered light, and $12s$ to solve for the background model Taylor expansion . The computational complexity of applying the matched filter scales with the number of filters applied, and with the size of each filter. As an example, it takes about 1.8 seconds to apply a matched filter of size $(25, 100)$ to $1116$ frames of $2048 \times 2048$ resolution on a single NVIDIA RTX 3090 GPU. The algorithm scales linearly with the number of frames, area of matched filter, and number of matched filters. Larger matched filters (required for closer objects) take longer to run.

The primary limitations of this method are that it cannot find objects that are extremely dim and far away (which unfortunately is what Planet Nine is estimated to be). This is most likely due to the problem of self-subtraction, because the further away an object, the smaller the effect of parallax and therefore distance traveled. This is a fundamentally hard problem, and the solution may involve further constraining the Taylor expansion parameters of the background model.

\subsection{Future Directions}

We plan to apply this method to the full TESS data set to extract all the
candidates for all sectors. We can then further reduce the number of false
positives by requiring candidates to occur in multiple observations in
different sectors. This will allow us to use a possibly higher threshold for
detection in any given sector thereby increasing the recall while also reducing
false positives. We will publish the detections online to enable others to
confirm the candidates via other observations.

\section{Conclusions}
\label{section:conclusion}

\section{Acknowledgments}
\label{section:acknowledgements}

We thank Elizabeth and George Ricker and Tansu Daylan for their encouragement, guidance and help along the way. This research has made use of data and/or services provided by the International Astronomical Union's Minor Planet Center and Mikulski Archive for Space Telescopes (MAST).\footnote{\url{http://archive.stsci.edu/tess/bulk_downloads/bulk_downloads_ffi-tp-lc-dv.html}} as well as TESS at NASA.

\software{\texttt{numpy} \citep{oliphant2006guide, walt2011numpy, harris2020array}, \texttt{matplotlib} \citep{hunter2007matplotlib}, \texttt{lightkurve} \citep{lightkurve2018}, \texttt{astroquery} \citep{ginsburg2019astroquery}, \texttt{PyEphem} \citep{rhodes2011pyephem}}, \texttt{astropy} \citep{astropy2013, astropy2018}, \texttt{scipy} \citep{virtanen2020scipy}, \texttt{pytorch} \citep{pytorch}

\bibliography{bibliography}
\bibliographystyle{aasjournal}

\end{document}